\documentclass[twocolumn,prc,aps,showpacs,amsmath,amssymb,nofootinbib]{revtex4}
\usepackage{graphicx}
 \newcommand\la{\langle}
 \newcommand\ra{\rangle}
 \newcommand\beq{\begin{equation}}
 
 \newcommand\eeq{\end{equation}}
 \newcommand\beqn{\begin{eqnarray}}
 \newcommand\eeqn{\end{eqnarray}}
 \newcommand\GeV{{\rm GeV}}
 
\def\im{\mbox{Im}\,}

\def\fm{\,\mbox{fm}}
\def\GeV{\,\mbox{GeV}}

\def\lsim{\mathrel{\rlap{\lower4pt\hbox{\hskip1pt$\sim$}}
    \raise1pt\hbox{$<$}}}         
\def\gsim{\mathrel{\rlap{\lower4pt\hbox{\hskip1pt$\sim$}}
    \raise1pt\hbox{$>$}}}         

\def\la{\langle}
\def\ra{\rangle}

\begin{document}

\title{Nuclear effects in leading neutron production}

\author{B. Z. Kopeliovich}
\author{I. K. Potashnikova}
\author{Iv\'an Schmidt}

\affiliation{\centerline{Departamento de F\'{\i}sica,
Universidad T\'ecnica Federico Santa Mar\'{\i}a; and}
Centro Cient\'ifico-Tecnol\'ogico de Valpara\'iso;
Casilla 110-V, Valpara\'iso, Chile}

\begin{abstract}
Absorptive corrections, known to  suppress proton-neutron transitions with large fractional momentum
$z\to1$ in $pp$ collisions,  become dramatically strong on a nuclear target, and push the partial cross sections of leading neutron production to the very periphery of the nucleus.
The mechanism of $\pi$-$a_1$ interference, which successfully explains the observed single-spin asymmetry in polarized  $pp\to nX$, is extended to collisions of polarized protons with nuclei. Corrected for nuclear effects, it explains the observed single-spin azimuthal asymmetry of neutrons, produced in inelastic events, where the nucleus violently breaks up. The single-spin asymmetry is found to be negative and nearly $A$-independent.

\end{abstract}



\maketitle

\section{Why neutron production?}

\begin{figure}[htb]
\centerline{
  \scalebox{0.3}{\includegraphics{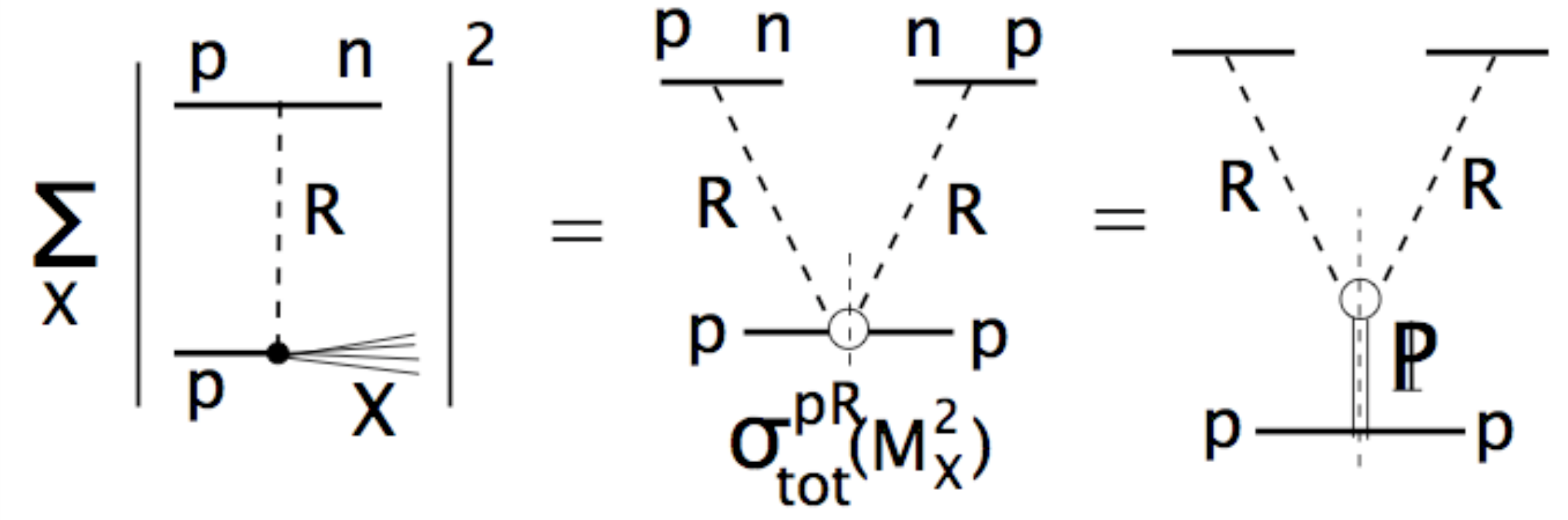}}}
\caption{\label{fig:3R} Graphical relation between 
the cross section of neutron production and the total Reggeon-proton cross section, which at large $M_X^2$ is dominated by the Pomeron.}
 \end{figure}

The process $p+p(A)\to n+X$,
with a large fractional light-cone momentum $z=p^+_n/p^+_p$
of neutrons produced in the proton beam direction, is known to
be related to the iso-vector Reggeons ($\pi,\ \rho,\ a_2,\ a_1$, etc.)
\cite{kpss},
as is illustrated in figure~\ref{fig:3R}, where
the amplitude, squared and summed over the final states $X$ (at a fixed invariant mass $M_X$), is expressed via the Reggeon-proton total cross section at the c.m. energy $M_X$. 

At  high energies of colliders (RHIC, LHC) $M_X^2=s(1-z)$ is so large (except  of the inaccessibly small $1-z$), that the cross section $\sigma^{pR}_{tot}(M_X^2)$ is dominated by the Pomeron exchange, as is illustrated in figure~\ref{fig:3R}. 

The couplings of iso-vector Reggeons with natural parity ($\rho$, $a_2$) to the proton  are known to be predominantly spin-flip \cite{kane}, so they can be neglected, because we are interested here in small transverse momenta of neutrons, $p_T\to0$. 
Only unnatural parity Reggeons ($\pi$, $a_1$), having large spin non-flip couplings contribute in the forward direction.

\section{Proton-to-neutron transition in the vicinity of pion pole}
 
Pions are known to have a large coupling with nucleons, so the
pion exchange is important in the processes with isospin flip, like  $p\to n$. 
Measurements with polarized proton beams supply more detailed information about the interaction dynamic.

The pion term in the cross section of neutron production reads \cite{kpss},
\beq
\left.z\,\frac{d\sigma^B(pp\to nX)}{dz\,dq_T^2}\right|_{\pi}=
f_{\pi/p}(z,q_T,q_L)\,
\sigma^{\pi^+ p}_{tot}(s'),
\label{100}
 \eeq
where $s'=M_X^2$; and superscript $B$ means that this is the Born approximation, ignoring absorptive corrections. 
$f_{\pi/p}(z,q_T,q_L)$ is the pion flux in the proton, having the form,
 \beqn
f_{\pi/p}(z,q_T,q_L)&=&
|t|\,G_{\pi^+pn}^2(t)\left|\eta_\pi(t)\right|^2
\left(\frac{\alpha_\pi^\prime}{8}\right)^2 
\nonumber\\ &\times&
(1-z)^{1-2\alpha_\pi(t)}.
\label{120}
 \eeqn
Here $q_T$ is the neutron transverse momentum; 
$
q_L=m_N(1-z)/\sqrt{z}
$; 
$
-t=q_L^2+q_T^2/z
$;  and $G_{\pi^+pn}(t)$ is the effective $\pi$-$N$ vertex function \cite{kpp,kpss}.

The amplitude of the process includes both non-flip and spin-flip terms \cite{kpss,kpss-spin},
\beq
A^B_{p\to n}(\vec q,z)=
\bar\xi_n\left[\sigma_3\, q_L+
\frac{1}{\sqrt{z}}\,
\vec\sigma\cdot\vec q_T\right]\xi_p\,
\phi^B(q_T,z),
\label{900}
\eeq

The procedure of inclusion of the absorptive corrections on the amplitude level was developed in \cite{kpss,kpps}. First, the amplitude (\ref{900}) is Fourier transformed to impact parameter representation, where the absorptive effect is just a multiplicative suppression factor. Then the absorption corrected amplitude is Fourier transformed back to momentum representation. The effects of absorption turn out to be quite strong, it about twice reduces the neutron production cross section, and affects 
differently the non-flip and spin-flip terms in the amplitude.
Here we skip the derivation, which is too lengthy and is described in  detail in \cite{kpss,kpps,pi-pi}.  

\section{Leading neutrons from \MakeLowercase{p}A collisions}

Being close to the pion pole, we treat the vertex $\pi+p\to X$ in figure~\ref{fig:3R} as the pion-proton interaction amplitude.
Squaring it, we get in (\ref{100}) the pion-proton total cross section.
In the case of a nuclear target we should replace the proton by a nucleus and get $\sigma_{tot}^{\pi A}(M_X^2)$.
The latter gets contributions from different channels, which can be classified as inelastic and diffractive interactions.
The former corresponds to multiparticles production filling the rapidity interval between the colliding pion and nucleus;
while the latter corresponds to rapidity gap events.
The corresponding cross sections can be evaluated within models as is described below.

The  recent measurements of forward neutrons in the PHENIX experiment was supplemented by the Beam-Beam Counters (BBC), detecting charged particles in two pseudo-rapidity intervals $3.0<|\eta|<3.9$. The results of measurements of forward neutrons are
presented for different samples of events:\\ (i) inclusive neutron production, with BBC switched off.\\
(ii) Neutrons accompanied with multi-particle production (one or both BBCs are fired), can be associated with inelastic
$\pi$-$A$ collisions;\\
(iii) If both BBC are vetoed, a large contribution of diffractive interactions might be expected.

Notice that such a correlation with BBC activities and related processes should not be taken literally, and comparison with theoretical predictions should be done with precaution. Further experimental studies employing Monte-Carlo simulations are required. 

\subsection{Glauber model}

A natural extension of equation (\ref{100}) to nuclear targets has the form,
\beq
\frac{d\sigma(pA\to nX)}{d\ln(z)dq_T^2d^2b_A}=
f_{\pi/p}(z,q_T)\frac{d\sigma^{\pi A}_{tot}(M_X^2)}{d^2b_A}\,
 S_{NA}(b_A),
\label{580}
\eeq
where $b_A$ is the impact parameter of $pA$ collision; $S_{NA}(b_A)$ is the additional nuclear absorption factor described below.

This expression can be interpreted as interaction of the projectile Fock component $|\pi^+n\ra$  with the target, sharing the proton light-cone momentum in fractions $z$ and $(1-z)$ respectively. While the pion interacts inelastically with the target, the spectator neutron has to remain intact, i.e. has to survive propagation through the nucleus.

The partial total pion-nucleus cross section in equation~(\ref{580}) can be evaluated in the Glauber approximation,
\beqn
\left.\frac{\sigma^{\pi A}_{tot}(M_X^2)}{d^2b_A}\right|_{Gl} 
\approx
2\left[1-
e^{-{1\over2}\,\sigma^{\pi N}_{tot}(M_X^2)\,
T_A(b_A)}\right]\, ,
\label{620}
\eeqn
 where 
$
T_A(b_A) = \int_{-\infty}^\infty d\zeta\,\rho_A(b_A,\zeta)\ ,
$
 is the nuclear thickness function; $\rho_A(b_A,\zeta)$
 is the nuclear density. Notice that for numerical calculations we use here and in what follows more 
 accurate form, replacing $e^{-{1\over2}\,\sigma^{\pi N}_{tot}T_A} \Rightarrow [1-\sigma^{\pi N}_{tot}T_A/2A]^A$.

Correspondingly, the neutron survival factor in (\ref{580}) reads,
\beqn
S_{NA}(b_A)\bigr|_{Gl} &\approx&
\frac{e^{-\sigma^{nN}_{in}(zs)T_A(b_A)}-e^{-\sigma^{pN}_{in}(s)T_A(b_A)}}
{T_A(b_A)[\sigma^{pN}_{in}(s)-\sigma^{nN}_{in}(zs)]}
\nonumber\\ &\approx& 
e^{-\sigma^{NN}_{in}(s)T_A(b_A)}.
\label{800}
\eeqn

If BBC are vetoed, the nucleus quite probably remains intact or decays into fragments without particle production. 
Then instead of the total $\pi A$ cross section one should use in (\ref{580})
diffractive cross section related to elastic, $\pi A\to\pi A$, and quasielastic $\pi A\to\pi A^*$, channels.
The corresponding cross section has the form  \cite{mine,kps},
\beqn
\left.\frac{\sigma^{\pi A}_{diff}(M_X^2)}{d^2b_A}\right|_{Gl} &=&
\left[1-e^{-{1\over2}\sigma^{\pi N}_{tot}T_A(b_A)}\right]^2
\nonumber\\ &+&
\sigma^{\pi N}_{el}T_A(b_A)e^{-\sigma^{\pi N}_{in}T_A(b_A)},
\label{890}
\eeqn
where the first and second terms correspond to  elastic and quasi-elastic scattering respectively.

The difference between the cross sections, equations~(\ref{580}) and (\ref{890}) is related to 
inelastic $\pi$-$A$ interactions, leading to multi-particle production. Correspondingly, one should modify equation (\ref{580}) replacing the total by inelastic $\pi A$ cross section \cite{mine,kps},
\beq
\left.\frac{\sigma^{\pi A}_{in}(M_X^2)}{d^2b_A}\right|_{Gl} =
1-e^{-\sigma^{\pi N}_{in}T_A(b_A)}
\label{885}
\eeq

\subsection{Gribov corrections: color transparency}

It is well known that the Glauber approximation is subject to Gribov inelastic shadowing corrections \cite{gribov69},
which are known to make the nuclear matter more transparent for hadrons \cite{zkl,mine} and affect both factors in (\ref{580}), suppressing $\sigma^{\pi A}_{tot}$ and increasing $S_{NA}(b_A)$.
We calculate the Gribov corrections to all orders of multiple interactions by employing the dipole representation, as is described in  \cite{zkl,mine,kps,gribov85}. 

Hadron wave function on the light front can be expanded over different Fock states, consisted of parton ensembles with various transverse positions $\vec r_i$ of the partons. The interaction cross section of such a hadron is averaged over the Fock states, $\sigma_{tot}^{hp}=\la\sigma(\vec r_i)\ra_h$, where $\sigma(\vec r_i)$ is the cross section of interaction
of the partonic ensemble with transverse coordinates $\vec r_i$ with the proton target.

Notice that high-energy partonic ensembles are eigenstates of interaction \cite{zkl}, i.e. the parton coordinates $\vec r_i$ remain unchanged during the interaction. Therefore the eikonal approximation employed in the Glauber model, Eqs.~(\ref{620})-(\ref{890}) should not be used for hadrons, but for their Fock components, and then the whole exponential terms
should be averaged \cite{zkl,gribov85}. This corresponds to the following replacements in Eqs.~(\ref{620})-(\ref{890}),
\beq
e^{-{1\over2}\sigma^{h N}_{tot}T_A}=
e^{-{1\over2}\left\la\sigma(\vec r_i)\right\ra_h\,T_A}
\Rightarrow
\left\la e^{-{1\over2}\sigma(\vec r_i)\,T_A}\right\ra_h.
\label{895}
\eeq
The difference between these averaging procedures is exactly the Gribov corrections \cite{zkl,gribov85}.

The result of averaging in (\ref{895}) for proton-nucleus interactions was calculated in \cite{kps} with a realistic saturated paramentrization of the dipole cross section \cite{kst2} and the quark-diquark model for the nucleon wave function. 
\beqn
\left\la e^{-{1\over2}
\sigma(r_T)T_A}\right\ra
=e^{-\frac{1}{2}
\sigma_0\,T_A(b)}
\sum\limits_{n=0}^\infty
\frac{[\sigma_0\,T_A(b)]^n}
{2^n\,(1+n\,\delta)\,n!},
\label{370}
 \eeqn
where 
\beq
\sigma_0(s)=\sigma_{tot}^{pp}(s)\left[1+
\frac{3R_0^2(s)}{8\la r_{ch}^2\ra_p}\right].
\label{375}
\eeq
We use $\la r_{ch}^2\ra_p=0.8\fm^2$ \cite{r-ch}, and energy dependent saturation radius
 $R_0(s)=0.88\,fm\,(s_0/s)^{0.14}$ with $s_0=1000\,GeV^2$  \cite{kst2}. 

Gluon shadowing corrections, corresponding to the triple-Pomeron term in diffraction,  were introduced as well, as is described in \cite{mine,kps}.

\begin{figure}[htb]
\centerline{
    \scalebox{0.3}{\includegraphics{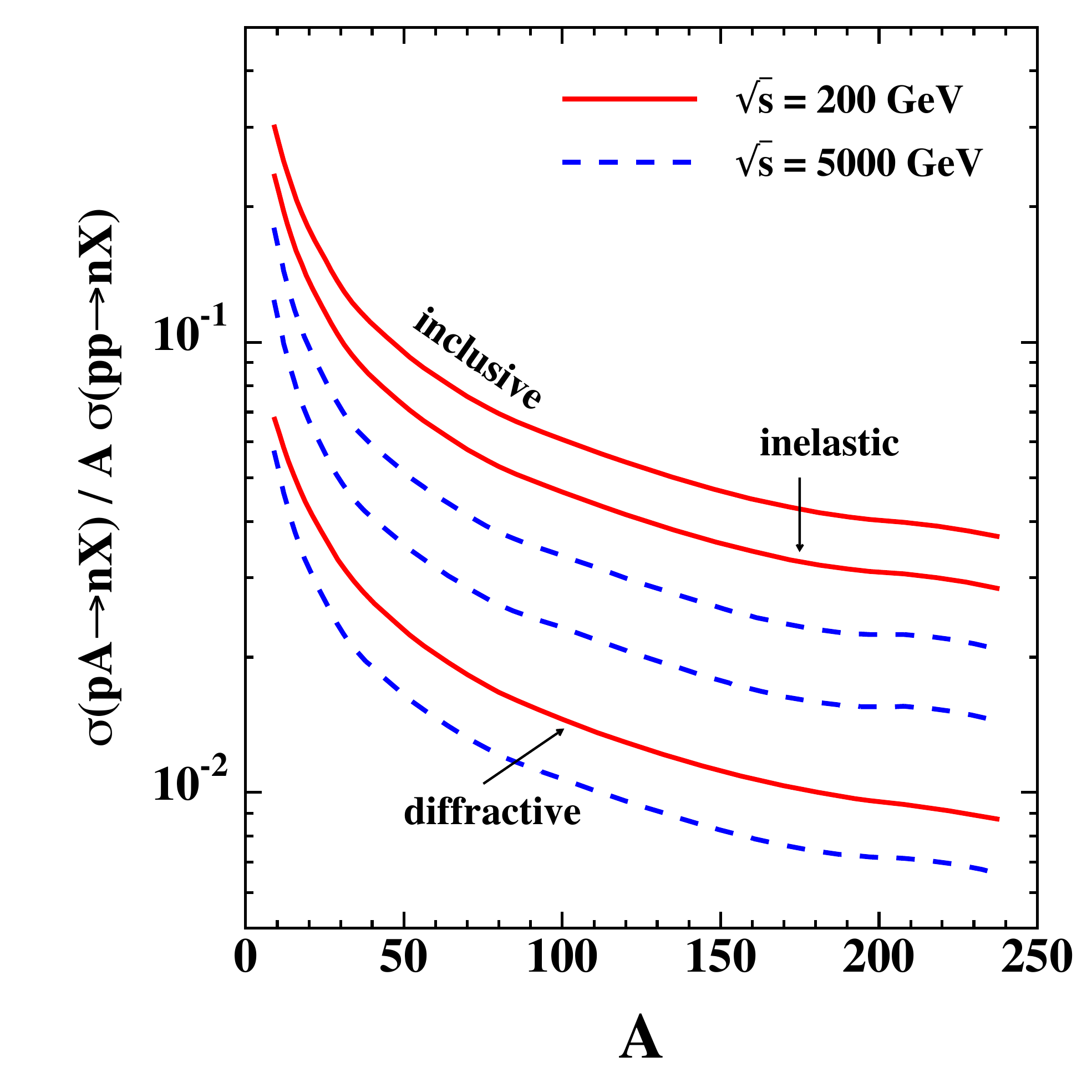}}
    }
\caption{\label{fig:sig2} $b_A$-integrated cross sections, normalized by the $pp$ cross section of leading neutron production. The solid and dashed curves correspond to both $pAu$ and $pPb$ collisions at $\sqrt{s}=200\GeV$ and $5000\GeV$ respectively. The three curves in each set from top to bottom correspond to inclusive, inelastic and diffractive neutron production, respectively.}
 \end{figure}

\begin{figure}[htb]
\centerline{
  \scalebox{0.3}{\includegraphics{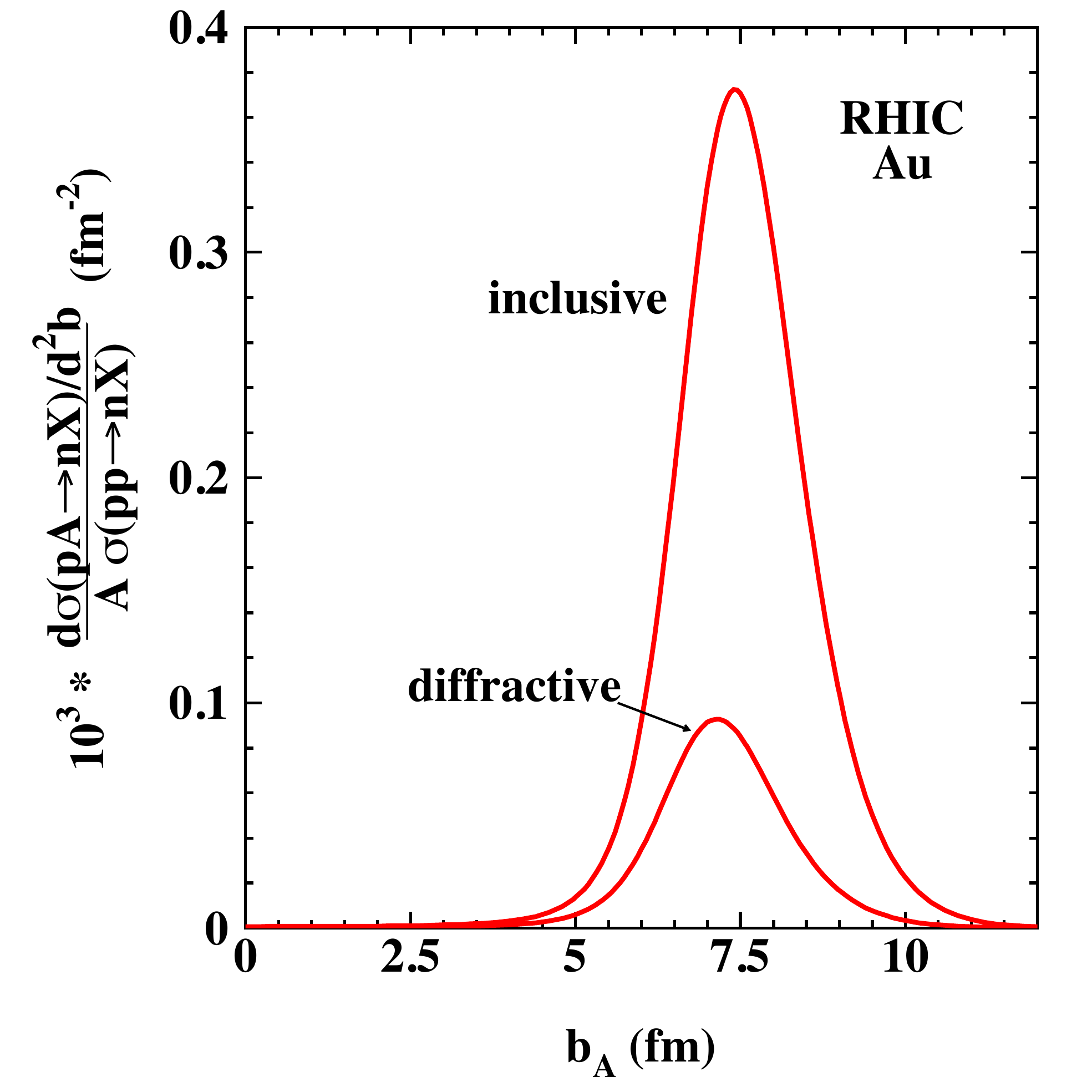}}
    }
\caption{\label{fig:sig1} Partial cross sections for inclusive (upper curve) and diffractive (bottom curve) neutron production in $pAu$ collisions at $\sqrt{s}=200\GeV$ and $z=0.75$.}
 \end{figure}

The results of Glauber model calculations, including Gribov corrections, for the partial inclusive cross section of $pAu\to nX$, equation~(\ref{580}),  normalized by the $pp\to nX$ cross section, are plotted in figure~\ref{fig:sig2} by the top solid red (RHIC) and top blue dashed (LHC) curves.
The two other lower curves show the cross section of inelastic and diffractive channels.
We see that the cross section is very small, what can be understood as a consequence of
significant suppression by the factor $S_{NA}(b_A)$ equation~(\ref{800}). Indeed, the impact parameter dependences of the 
inclusive, equation~(\ref{620}), and diffractive, equation~(\ref{890}), cross sections of neutron production, are depicted in figure~\ref{fig:sig1}. 
One can see that neutrons are produced from the very periphery of the nucleus, this is why the $b_A$-integrated cross section is so small. Correspondingly, the ratio shown in figure~\ref{fig:sig2} is falling for heavy nuclei as $A^{-2/3}$.

\section{Single-spin asymmetry}
\subsection{\boldmath$A_N$ in polarized $pp\to nX$}

Both terms in the amplitude equation~(\ref{900}) have the same phase factor, $\eta_\pi(t)=i-ctg\left[\pi\alpha_\pi(t)/2\right]$.
Therefore, in spite of the presence of both spin-flip and non-flip amplitudes, no single-spin asymmetry associated with pion exchange is possible in the Born approximation. Even the inclusion of absorptive corrections leave the spin effects miserably small \cite{kpss-trieste,kpss-spin} compared to data \cite{phenix-pp1,phenix-pp2}. 

A plausible candidate to generate a sizeable spin asymmetry at high energies is $a_1$ meson exchange,
since $a_1$ can be produced by pions diffractively. However, this axial-vector resonance 
is hardly visible in diffractive channels $\pi+p\to 3\pi+p$, which are dominated by $\pi\rho$
in the $1^+S$ wave. The $\pi\rho$ invariant mass distribution forms a pronounced narrow peak at $M_{\pi\rho}\approx m_{a_1}$ (due to the Deck effect \cite{deck}). Although in the dispersion relation for the amplitude this channel corresponds to a cut, it can be replaced with good accuracy by an effective pole $\tilde a_1$ \cite{belkov,kpss-spin}.
In the crossed channel, $\pi\rho$ exchange corresponds to a Regge cut, with known intercept and slope of the Regge trajectory \cite{kpss-spin}.

The expression for the single-spin asymmetry arising from $\pi$-$\tilde a_1$ interference has the form \cite{kpss-spin},
\beqn
&&A_N^{(\pi-\tilde a_1)}(q_T,z) =
q_T\,\frac{4m_N\,q_L}{|t|^{3/2}}\,
(1-z)^{\alpha_\pi(t)-\alpha_{\tilde a_1}(t)}
\label{920}
\\ &\times&
\frac{\im\,\eta_\pi^*(t)\,\eta_{\tilde a_1}(t)}
{\left|\eta_\pi(t)\right|^2}\,
\left(\frac{d\sigma_{\pi p\to \tilde a_1p}(M_X^2)/dt|_{t=0}}
{d\sigma_{\pi p\to\pi p}(M_X^2)/dt|_{t=0}}\right)^{1/2}
\frac{g_{\tilde a_1^+pn}}{g_{\pi^+pn}}.
\nonumber
\eeqn
The trajectory of the $\pi\rho$ Regge cut and the phase factor $\eta_{\tilde a_1}(t)$ are known from Regge phenomenology.
The ${\tilde a_1}NN$ coupling was evaluated in \cite{kpss-spin}, based on PCAC and the 
second Weinberg sum rule, where the spectral functions of the vector and axial
currents are represented by the $\rho$ and the effective ${\tilde a_1}$ poles respectively.
This leads to the following relations between the couplings,
\beq
\frac{g_{\tilde a_1 NN}}{g_{\pi NN}}=
\frac{m_{\tilde a_1}^2\,f_\pi}{2m_N\,f_\rho}\approx {1\over2},
\label{340}
\eeq
where $f_\pi=0.93m_\pi$ is the pion decay coupling;
$f_\rho=\sqrt{2}m_\rho^2/\gamma_\rho$, 
and $\gamma_\rho$ is the universal coupling, $\gamma_\rho^2/4\pi=2.4$.

The parameter-free calculations of $A_N$ in $pp\to nX$ \cite{kpss-spin} agree well with the PHENIX data \cite{phenix-pp1,phenix-pp2}.

\subsection{\boldmath$A_N$ in polarized $pA\to nX$}

The single-spin asymmetry on a nuclear target due to $\pi$-$\tilde a_1$ interference can be calculated with a modified equation (\ref{920}), in which one should replace  
\beq
\frac{d\sigma_{\pi p\to \tilde a_1p}(M_X^2)/dt|_{t=0}}
{d\sigma_{\pi p\to\pi p}(M_X^2)/dt|_{t=0}}
\Rightarrow
\frac{d\sigma_{\pi A\to \tilde a_1A}(M_X^2)/dt|_{t=0}}
{d\sigma_{\pi A\to\pi A}(M_X^2)/dt|_{t=0}}
\label{500}
\eeq
This replacement leads to the single-spin asymmetry, which can be presented in the form,
\beq
A_N^{pA\to nX} =
A_N^{pp\to nX}\times \frac{R_1}{R_2}\,R_3.
\label{520}
\eeq 

Factor $R_1$ in accordance with (\ref{920}) and (\ref{500}) is the nuclear modification factor for the forward amplitude of $\pi A\to \tilde a_1 A$ coherent diffractive transition. In the Glauber approximation it has the form,
\beqn
&&R_1=\int d^2b_A\int\limits_{-\infty}^\infty d\zeta\,
\rho_A(b_A,\zeta) 
\nonumber\\ &\times&
\exp\left[-{1\over2}\sigma^{\pi N}_{tot} T_-(b_A,\zeta)
-{1\over2}\sigma^{\tilde a_1 N}_{tot} T_+(b_A,\zeta)\right]\!,
\label{525}
\eeqn
where $T_-(b_A,\zeta)=\int_{-\infty}^\zeta d\zeta^\prime\, \rho_A(b_A,\zeta^\prime)$ and $T_+(b_A,\zeta)=
T_A(b_A)-T_-(b_A,\zeta)$.

Integrating over $\zeta$ analytically, we arrive at,
\beqn
R_1&=& \frac{1}{\Delta\sigma}\int d^2b_A\, e^{-{1\over2}\sigma^{\pi p}_{tot}T_A(b_A)}
 \nonumber\\ &\times&
\left[1-e^{-{1\over2}\Delta\sigma T_A(b_A)}\right]
 e^{-{1\over2}\sigma^{pp}_{tot}T_A(b_A)},
 \label{540}
 \eeqn
 where $\Delta\sigma=\sigma^{\tilde a_1 N}_{tot}-\sigma^{\pi N}_{tot}$.
 As was mentioned above and motivated in detail in \cite{kpss-spin,belkov,pcac}, diffractive
production of $a_1$ axial-vector meson is a very weak signal compared with $\rho$-$\pi$ production, which form a rather narrow peak in the invariant mass distribution.
Therefore  with a good accuracy $\sigma^{\tilde a_1 N}_{tot}=\sigma^{\rho N}_{tot}+\sigma^{\pi N}_{tot}$, and
$\Delta\sigma=\sigma^{\rho   N}_{tot}$. Data for photoproduction of $\rho$ meson on nuclei agree with
$\sigma^{\rho   N}_{tot}\approx \sigma^{\pi   N}_{tot}$, so we fix $\Delta\sigma$ at this value.

 Nuclear modification, corresponding to the denominator of equation~(\ref{500}) is determined by
Eqs.~(\ref{580}) - (\ref{800}) and has the form,
 \beqn
 R_{2}= \frac{2}{\sigma^{\pi p}_{tot}}\int d^2b_A
\left[1-e^{-{1\over2}\sigma^{\pi p}_{tot}T_A(b_A)}\right]
 e^{-{1\over2}\sigma^{pp}_{tot}T_A(b_A)}.
 \label{560}
 \eeqn
  
\begin{figure}[htb]
\centerline{
 {\includegraphics[height=7cm]{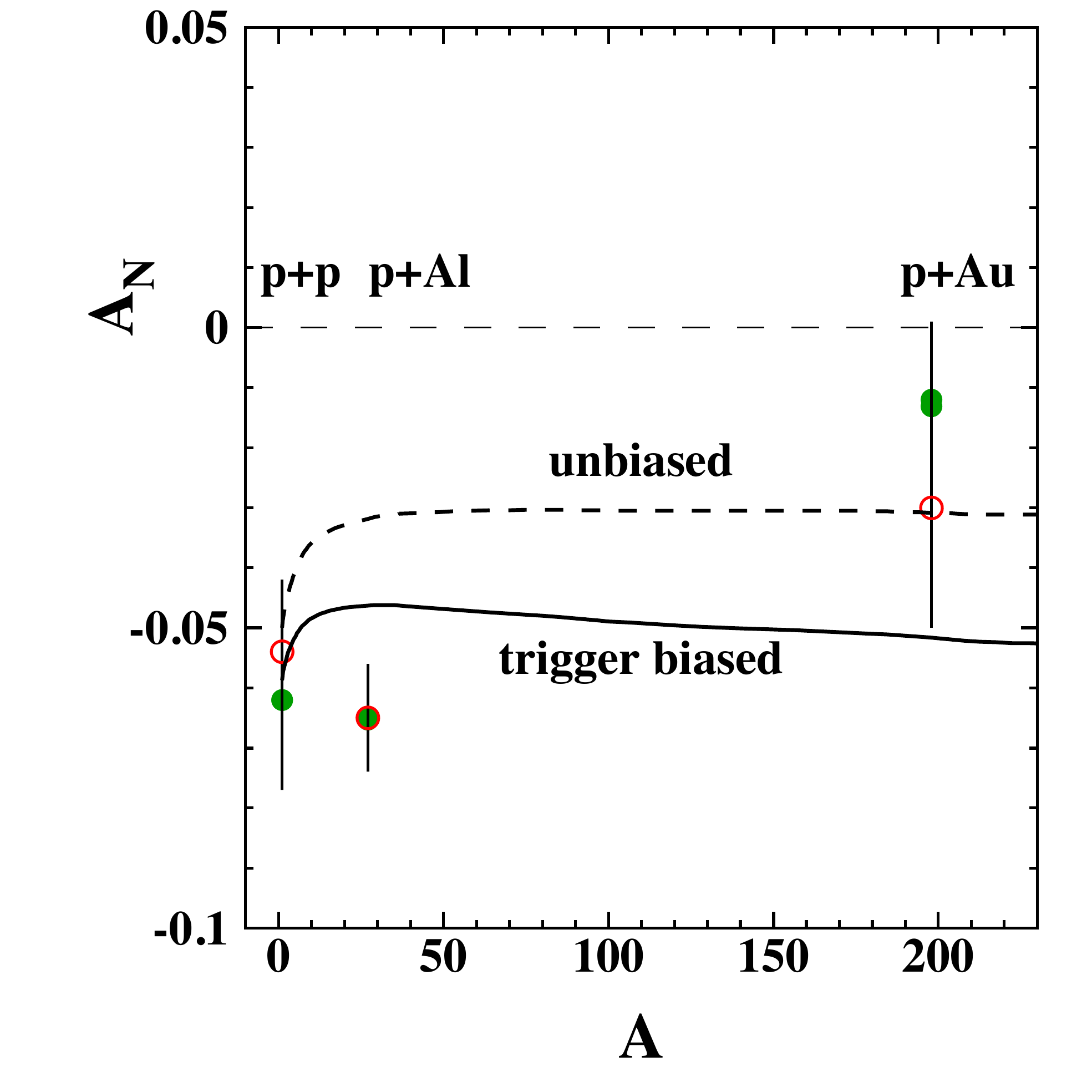}}
 }
\caption{\label{fig:AN} $A_N$ in polarized $pA\to nX$ vs $A$ at $\sqrt{s}=200\GeV$, $\la q_T\ra=0.115\GeV$ and $\la z\ra = 0.75$. Full and open data points correspond to events with either both BBCs fired, or only one of them fired in the nuclear direction, respectively \cite{bazil1,itaru,bazil2}. An attempt to model these two classes of events is presented by solid and dashed curves  (see text).
}
 \end{figure}

The factor $R_3$ depends on how the measurements were done.
If the BBC are fired, a proper estimate would be 
$R_3=\sigma^{\pi A}_{tot}/\sigma^{\pi A}_{in}$. Otherwise, if the BBC are switched off (inclusive neutron productions), we fix $R_3=1$. The results corresponding to these two choices are plotted in figure~\ref{fig:AN}, by solid and dotted curves respectively.
All data points correspond to events with BBC fired. However full green and open red points correspond to events with either both BBCs fired, or only one of them in the nuclear direction, respectively \cite{bazil1,itaru,bazil2}.

The difference between these two results reflects the uncertainly in the physical interpretation of events with fired of vetoed BBCs. This can be improved by applying a detailed Monte-Carlo modelling. Nevertheless, the results of our calculations,
presented in figure~\ref{fig:AN}, reproduce  reasonably well experimental data \cite{bazil1,itaru,bazil2}.

\begin{figure}[htb]
\centerline{
  \scalebox{0.3}{\includegraphics{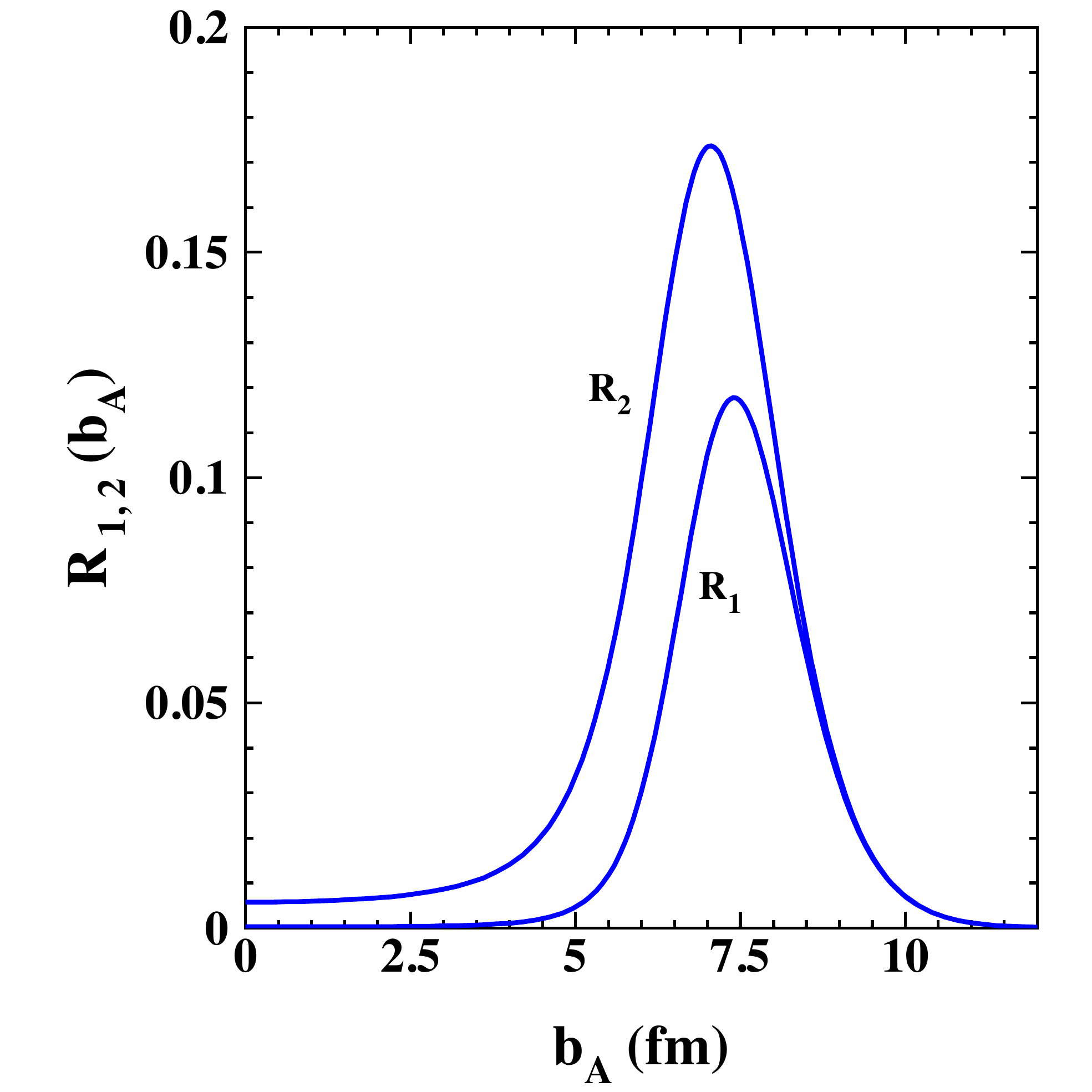}}
    }
\caption{\label{fig:R12} Partial cross sections for inclusive (upper curve) and diffractive (bottom curve) neutron production in $pAu$ collisions at $\sqrt{s}=200\GeV$.}
 \end{figure}

A remarkable feature of the single-spin asymmetry $A_N$ of neutrons produced on nuclear targets is a very weak $A$-dependence, seen both in the data and our calculations. The reason can be easily understood. All $A$-dependence of
the asymmetry $A_N$ is contained in the factors $R_1$ and $R_2$ in equation~(\ref{520}). It turns out that the strong nuclear absorption factor $S_{NA}(b_A)$, equation~(\ref{800}), contained in both equations (\ref{540}) and (\ref{560}), push
neutron production to the very periphery of the nucleus. This is demonstrated by the $b_A$-unintegrated factors $R_1(b_A)$ and $R_2(b_A)$, plotted  in figure~\ref{fig:R12} for gold at $\sqrt{s}=200\GeV$.
Due to the observed similarity of $A$-dependences of $R_1$ and $R_2$, $\propto A^{1/3}$, they mostly cancel in (\ref{520}), resulting in nearly $A$-independent single-spin asymmetry of neutrons.

\section{Summary}

The previously developed methods of calculation of the cross section of leading neutron production in $pp$ collisions, are extended to nuclear targets. The nuclear absorptive corrections, calculated in the Glauber-Gribov approach, are so strong that push the partial cross sections of leading neutron production to the very periphery of the nucleus. As a result, the $A$-dependences of inclusive and diffractive neutron production turn out to be similar, $\sim A^{1/3}$. 

The mechanism of $\pi$-$a_1$ interference, which successfully explains the observed single-spin asymmetry in polarized reaction $pp\to nX$, is extended to collisions of polarized protons with nuclei. Corrected for nuclear effects, it explains quite well the observed asymmetry $A_N$ in inelastic events, when the nucleus violently breaks up \cite{bazil1,itaru,bazil2}. However, the large value and opposite sign of $A_N$ observed in the diffractive sample, still remains a challenge.\\

\begin{acknowledgments}
We are thankful to Alexander Bazilevsky, Itaru Nakagawa and Minjung Kim 
for providing us with preliminary data and details of the measurements, as well as
for numerous informative discussions.
This work was supported in part
by Fondecyt (Chile) grants 1130543, 1130549 and 1140842,
by Proyecto Basal FB 0821 (Chile),
and by CONICYT grant  PIA ACT1406 (Chile).
\end{acknowledgments}

\end{document}